\documentstyle[12pt,epsfig]{article}
\begin{document}
\title{Mass of the Lowest Scalar $(0^{++})$ Glueball in the QCD Sum Rules}
\author{{Tao Huang and Ai-lin Zhang}\\
{CCAST (World Laboratory), P. O. Box 8730, Beijing, 100080}\\ {and}\\
{Institute of High Energy Physics, P. O. Box 918, Beijing, 100039}\\ {P. R.
China}}
\date {}
\maketitle
\begin{center}
\begin{abstract}
The mass of the lowest scalar glueball is discussed by using QCD sum rules. We find
that the glueball mass is sensitive to the choice of moments and
slightly depends on the radiative corrections. With the help of suitable moments
and stability criteria, we get the scalar glueball mass: $1710\pm110$ MeV without
radiative corrections and $1580\pm150$ MeV with radiative corrections.
\end{abstract}
\end{center}
\section{Intoduction}
\indent
\par
The existence of bound gluon states, glueballs, is a direct consequence
based on the QCD self-interactions among gluons. Although there are several
glueball candidates experimentally, there is no conclusive evidence on them.
Recently, one pays particular attention to three states: $f_0$(1500)(J=0),
$f_J$(1710) (J=0 or 2), and $\xi$(2230) (J$\geq2$). They seem like glueball
candidates of $0^{++}$ or 2$^{++}$ states.\par The properties of glueballs
have been investigated in the lattice gauge theory and in many models based
on the QCD theory. Even in the lattice gauge calculation, there are
different predictions for the $0^{++}$ glueball. IBM group \cite{w} predicts
the lightest $0^{++}$ glueball mass, $(1740\pm71)$  MeV, and UK QCD
group \cite{ukqcd} gives the estimated mass, $(1550\pm50)$ MeV respectively. They
give the slightly diffrerent predictions for the $2^{++}$ glueball:
$(2259\pm128)$  MeV (IBM group)and $(2270\pm100)$  MeV (UK QCD group). The
mass of the $0^{++}$ glueball can not be predicted consistent at present.
It encourage us to restudy the mass of the lowest $0^{++}$ glueball using QCD sum rules in this
paper.

\par V. A. Novikov {\it et al} \cite{nsvz} first tried to estimate the scalar
glueball mass by using QCD sum rules, but they only took the mass to be
700 MeV by hand because of uncontrolled instanton contributions. Since then,
P. Pascual and R. Tarrach \cite{pt}, S. Narison \cite{n} and J. Bordes et
al \cite{bgp} presented their calculation on the scalar glueball mass in the
framework of QCD sum rules. They all got a lower mass prediction around
700 MeV-900 MeV when they neglected the radiative corrections in their
calculation of the correlator, Namely, they only considered the perturbative
and the leading condensates in the correlator. E. Bagan and T.
Steele \cite{bs} first took account of the radiative corrections in the
correlator calculation, they got the higher mass prediction around $1.7$ GeV
since the one-loop $\langle\alpha_sG^2\rangle$ correction played a important
role. It seems that the radiative corrections make a big difference on the
prediction of the scalar glueball mass.\par In order to calculate the
predicted mass for the lowest scalar glueball in the QCD sum rules, we
re-study the correlator without radiative corrections and correlator with
radiative corrections, respectively.
After Borel transformation of the correlator weighted by different powers of
$Q^2$, we get different moments. It is the moments we choose which make the main
difference of the glueball mass, We give the criteria to choose the
continuum threshold that represents the maximum energy for which a duality exists between resonance
physics and QCD, and give some comments on a reasonable choice of the moments
also.The radiative corrections shift the mass scale a little. Our predicted
mass for $0^{++}$ glueball is in agreement with the result of UK QCD group.
\par The paper is organized as follows. In sect. 2 a brief review
about the calculation of scalar glueball mass from QCD sum rules was given. In sect.
3 we give the criteria to choose $s_0$ and the moments, and present  numerical
results. Finally, the last section is reserved for a summary.
\section{Scalar glueball sum rules}
\indent
\par
Let us consider the correlator
\begin{equation}
\Pi(q^2)={\it i}\int e^{{\it i}qx}\langle0|T\{j(x),j(0)\}|0\rangle d x,        
\end{equation}
where the current $j(x)$ is defined as
\begin{equation}\label{current}
j(x)=\alpha_sG_{\mu\nu}^aG_{\mu\nu}^a(x).               
\end{equation}
$G_{\mu\nu}^a$ in Eq.(\ref{current}) stands for the gluon field strenth tensor and $\alpha_s$ is the
quark-gluon coupling constant. The current $j(x)$ is the gauge-invariant and
non-renormalization \cite{t} (to two loops order) scalar current for the
$0^{++}$ glueball in QCD without quarks. We will keep all of the calculations in QCD
without quarks. 
\par Through operator products expansion,  the
correlator without radiative corrections becomes
\begin{eqnarray}\label{pi}
\Pi(q^2)&=&a_0(Q^2)^2\ln(Q^2/\nu^2)+b_0\langle\alpha_sG^2\rangle\\\nonumber
&+&c_0\frac{\langle g
G^3\rangle}{Q^2}+d_0\frac{\langle\alpha_s^2G^4\rangle}{(Q^2)^2},
\end{eqnarray}
with $Q^2=-q^2>0$, and
\[
\begin{array}{lllllll}
a_0&=&-(\frac{\alpha_s}{\pi})^2&,&b_0&=&4\alpha_s,\\
c_0&=&8\alpha_s^2&,&d_0&=&8\pi\alpha_s.
\end{array}
\]
After taking into account radiative corrections, the correlator is
\begin{eqnarray}
\Pi(q^2)&=&(a_0+a_1\ln(Q^2/\nu^2))(Q^2)^2\ln(Q^2/\nu^2)\\\nonumber
&+&(b_0+b_1\ln(Q^2/\nu^2))\langle\alpha_sG^2\rangle\\\nonumber
&+&(c_0+c_1\ln(Q^2/\nu^2))\frac {\langle gG^3\rangle}{Q^2}+d_0\frac{\alpha_s^2G^4}{(Q^2)^2}.
\end{eqnarray}                                                                 
where
\begin{eqnarray*}
a_0&=&-2(\frac {\alpha_s}{\pi})^2(1+\frac {51}{4}\frac {\alpha_s}{\pi}),\\
b_0&=&4\alpha_s(1+\frac {49}{12}\frac {\alpha_s}{\pi}),
\end{eqnarray*}
\[
\begin{array}{lll}
c_0=8\alpha_s^2,&d_0=8\pi\alpha_s&,\\
a_1=\frac {11}{2}(\frac {\alpha_s}{\pi})^3,&b_1=-11\frac{\alpha_s^2}{\pi},&c_1=-58\alpha_s^3.
\end{array}
\]
For the non-perturbative condensates we use the following notations and
estimate
\begin{eqnarray*}
\langle\alpha_sG^2\rangle&=&\langle\alpha_sG_{\mu\nu}^aG_{\mu\nu}^a\rangle,\\
\langle g G^3\rangle&=&\langle g f_{a b c}G_{\mu\nu}^a G_{\nu\rho}^b
G_{\rho\mu}^c\rangle,\\
\langle\alpha_s^2G^4\rangle&=&14\langle(\alpha_s f_{a b c}
G_{\mu\rho}^a G_{\rho\nu})^2\rangle-\langle(\alpha_s f_{a b c} 
G_{\mu\nu}^a G_{\rho\lambda}^b)^2\rangle.
\end{eqnarray*}
Now,we can use the standard dispersion representation for the correlator
\begin{equation}
\Pi(Q^2)=\Pi(0)-\Pi^{\prime}(0)+\frac
{(Q^2)^2}{\pi}\int_{0}^{+\infty}\frac{Im\Pi(s)}{s^2(s+Q^2)}d s
\end{equation}
to connect the QCD calculation with the resonance physics. From the low energy theorem \cite{nsvz} follows that
\begin{equation}
\Pi(0)=\frac{32}{11}\pi\langle\alpha_sG^2\rangle .
\end{equation}
\par For the physical spectral density $Im\Pi(s)$, one can divide it into
two parts, low energy part and high energy part. Fortunately, its
high-energy behavior is known as trivial,
\begin{equation}
Im\Pi(s)\longrightarrow\frac{2}{\pi}s^2\alpha_s^2(s),           
\end{equation}
while at low energy, $Im(s)$ can be expressed in the
narrow width approximation. The single resonance model for $Im\Pi(s)$ leads
\begin{equation}
Im\Pi(s)=\pi f^2M^4\delta(s-M^2),                  
\end{equation}
where M, f are the mass and coupling of the state. With these in hand, we
can proceed the following calculation.
\section{Moments and numerical results}
\indent
\par
To construct the sum rules,we define the moments $R_k$,
\begin{eqnarray}\label{moment}
R_k(\tau, s_0)&=&\frac{1}{\tau}\hat{L}[(Q^2)^k\{\Pi(Q^2)-\Pi(0)\}]\\\nonumber
&-&\frac{1}{\pi}\int_{s_0}^{+\infty}s^k e^{-s\tau}Im\Pi^{\{pert\}}(s)d s ,  
\end{eqnarray}
where $\tau$ is the Borel transformation parameter, $\hat{L}$ is the Borel tranformation
operator and $s_0$ represents the maximum energy for which a duality exists
between resonance physics and QCD calculation with condensates.
\par The standard dispersion relation is transformed into
\begin{equation}
R_k(\tau, s_0)=\frac{1}{\pi}\int_{0}^{s_0}s^ke^{-s\tau}Im\Pi(s)ds,
\end{equation}
and from Eq.(\ref{moment}) we have (for $k\geq-1$ )
\begin{equation}
R_k(\tau, s_0)=(-\frac{\partial}{\partial\tau})^{k+1}R_{-1}(\tau,s_0).
\end{equation}
The moment $R_{-1}(\tau,s_0)$ with radiative corrections has been given\cite{bs},
\begin{eqnarray}\label{rm}
R_{-1}(\tau,s_0)&=&-\frac{a_0}{\tau^2}\lfloor1-\rho_1(s_0\tau)\rfloor\\\nonumber
&+&2\frac{a_1}{\tau^2}(\gamma_E+E_1(s_0\tau)+\ln s_0\tau+\exp(-s_0\tau)-1
\\\nonumber
&-&[1-\rho_1(s_0\tau)]\ln \frac{s_0}{\nu^2})+\Pi(0) \\\nonumber
&-&\{b_0-b_1[\gamma_E+\ln \tau\nu^2+E_1(s_0\tau)]\}\langle\alpha_sG^2\rangle
\\\nonumber
&-&[c_0+c_1(1-\gamma_E-\ln\tau\nu^2-E_1(s_0\tau)+\frac{\exp(-s_0\tau)}{s_0\tau})]
\langle gG^3\rangle \tau \\\nonumber
&-&\frac{1}{2}d_0\langle \alpha_s^2G^4\rangle\tau^2 ,  
\end{eqnarray}
where
\[
\begin{array}{lll}
\rho_1(s_0\tau)=(1+s_0\tau)e^{-s_0\tau}&,&E_1(x)=\int_{x}^{\infty}\frac{e^{-y}}{y}dy,
\end{array}
\]
and $\gamma_E=Euler's constant\approx 0.5772$.
Renormalization-group improvement of the sum rules amounts to the substitution:
\begin{eqnarray*}
\nu^2 &\rightarrow & \frac{1}{\tau}, \\
\langle g G^3\rangle & \rightarrow & [ \frac{\bar{\alpha_s}}
{\bar{\alpha_s}(\nu^2)} ]^{7/11}\langle g G^3\rangle.
\end{eqnarray*}
\par $R_{-1}(\tau,s_0)$ without radiative corrections can be obtained as the
coefficients $a_1, b_1$, and $c_1$ are set to zero and $a_0, b_0, c_0$ from
Eq. (\ref{pi}).
\par Complete knowledge of $\Pi(Q^2)$ would allow us to fix the mass and
width of the glueball, but we are far from this idea. One can only
choose some suitable moments at appropriate $s_0$ to derive the prediction.
As shown in Ref.\cite{bs}, the $R_{-1}$ sum rule leads to a much smaller mass
scale due to the anomalously large contribution of the low-energy part
$\Pi(0)$ of the sum rule and it violates asymptotic freedom at large energy.
They claimed that $R_{-1}$ is not reliable to predict the $0^{++}$ glueball
mass. They employed the $R_0$ and $R_1$ sum rules to predict the $0^{++}$
glueball mass by fitting the stability criteria with the radiative
corrections considered. Their approach shows that the $R_0$ and $R_1$ sum
rules with the radiative corrections can obtain a higher mass scale
compared to the previous approaches, thus one would ask: which
factor(to choose an appropriate moments $R_k$ or to consider the radiative
corrections )is crucial to get the higher mass prediction? In order to answer
it, we re-examine the $R_k$ sum rules.
\par To improve the convergence of the asymtotic series, we study the ratio $\frac{R_{k+1}}{R_k}$ , such as
$\frac{R_0}{R_{-1}}$ and $\frac{R_1}{R_0}$ .
In the narrow width approximation, we have
$$M^{2k+4}f^2\exp(-\tau M^2)=R_k(\tau,s_0),$$
and(with $k\geq-1$)
\begin{equation}\label{m}
M^2(\tau,s_0)=\frac{R_{k+1}(\tau,s_0)}{R_k}.
\end{equation}
To proceed calculation, we choose the following parameters
\begin{eqnarray*}
\langle\alpha_sG^2\rangle&=&0.06 GeV^4,\\
\langle gG^3\rangle&=&(0.27 GeV^2)\langle\alpha_sG^2\rangle,\\
\langle\alpha_s^2G^4\rangle&=&\frac{9}{16}\langle\alpha_sG^2\rangle^2,\\
\Lambda_{\bar{MS}}&=&200 MeV,\\
\bar{\alpha_s}&=&\frac {-4\pi}{11\ln(\tau\Lambda_{\bar{MS}}^2)}.
\end{eqnarray*}
\par $M^2$ and $f^2$ are the functions of $s_0$ which is the starting point of
the continuum threshold, $s_0>M^2$. Since the glueball mass $M$ in
Eq.(\ref{m}) depends on $\tau$ and $s_0$ , we take the stationary point of
$M^2$ versus $\tau$ at an appropriate $s_0$ as the square of the glueball mass.
\par To determine the appropriate $s_0$ , the following stability criteria
are employed: (1), $s_0$ should be a little higher than the physcical mass and
approches it as near as possible due to the continuum threshold hypothesis
and the narrow width approximation; (2), The choice of a suitable $s_0$ should
lead to not only a widest flat portions of the plots of $M^2$ versus $\tau$
but also an appropriate parameter region of $\tau$ with the parameter region
comparable to the value of the glueball mass.
\par First, we start the $R_k$ sum rules without radiative corrections
to see the influence of different moments. According to the criteria above,
the acceptable region of $s_0$ is chosen between $s_0=3.0$ GeV$^2$ and $s_0=4.3$
GeV$^2$. Let us start with the $R_{-1}$, the ratio
$\frac{R_0}{R_{-1}}$ results in a much smaller mass scale which can't be
comparable to the parameter region(see Fig. 1) and it is not acceptable.
This result is similar to that as pointed out in Ref.\cite{bs} .The numerical
results of $\frac{R_1}{R_0}$ and $\frac{R_2}{R_1}$ without radiative
corrections are illustrated in Fig. 2 and Fig. 3, respectively. In the
figures, the optimum parameter of $s_0$ is chosen as $s_0=3.6$ GeV$^2$. The ratio $\frac{R_1}{R_0}$
can get a higher mass scale, but the parameter region of the $\tau$ can't be
comparable to the mass scale(the parameters corresponding to the stationary
point are too low ), so it doesn't satisfy the criteria above. We
won't take it for the mass prediction. The ratio $\frac{R_2}{R_1}$ in Fig.
3 gives an excellent shape and it satisfies all of the
criteria. The curve shows that the $0^{++}$ glueball mass is $1710$ MeV. In
the acceptable region of $s_0$, the $0^{++}$ glueball mass is $1710\pm110$
GeV. The
moments with higher $k$ can't stress the resonance contribution in the sum
rule, and the higher dimension condensates will contribute to the sum
rule(but we learn little about higher dimension condensates at present), we
think they are unsuitable for the mass prediction either.
\par As the radiative corrections are taken into account, the predicted mass
from ratio
$\frac{R_2}{R_1}$ only shifts the glueball mass a little, and the values are a
little lower ($17\% \sim 9\%$) than those predicted from the $\frac{R_2}{R_1}$
without radiative corrections in the acceptable region of $s_0$. The curve in Fig. 4 shows that the glueball
mass is $1580$ MeV with radiative corrections. In
the acceptable region of $s_0$, the $0^{++}$ glueball mass is $1580\pm150$
MeV.  
\section{Summary}
\indent
\par In this paper we re-study the scalar glueball mass based on the duality
among resonance physics and QCD. The modified Borel transformation has been
employed, it make the calculation more convenient and reasonable.
\par We find that the predicted  mass is sensitive to the choice of the
moment, and it is the moment which makes different predictions. Not all the moments are suitable for predicting the glueball mass, the
moments $R_{-1}$ and $R_k$ with higher $k$ aren't suitable. The moment
$R_{-1}$, the contribution of
low energy part of the correlator is large; The moments with higher $k$, 
the contribution of higher dimension condensates  will come into play.
\par To stress the contribution of resonance in sum rules, we give our
criteria on the choice of the continuum threshold, these criteria also make
it possible to choose a suitable moment for the calculation of the glueball
mass. We conclude that the ratio $\frac{R_2}{R_1}$ is the best one for the
calculation.
\par The radiative corrections shift the mass a little which is lower than that
without radiative corrections. The  numerical results are obtained:
$1710\pm110$
MeV without radiative corrections, and $1580\pm150$ MeV with radiative
corrections. The predicted mass is in agreement with the result of UK QCD
group and consistent with the glueball candidate $f_0$(1500)(J=0).

\newpage
\par
{\bf Figure caption}
\par
Figure 1: $\frac{R_0}{R_{-1}}$ versus $\tau$ at $s_0=3.6$ GeV$^2$ without
radiative corrections.

\par
Figure 2: $\frac{R_1}{R_0}$ versus $\tau$ at $s_0=3.6$ GeV$^2$ without
radiative corrections.

\par
Figure 3: $\frac{R_2}{R_1}$ versus $\tau$ at $s_0=3.6$ GeV$^2$ without
radiative corrections.

\par
Figure 4: $\frac{R_2}{R_1}$ versus $\tau$ at $s_0=3.6$ GeV$^2$ with radiative
corrections.
\newpage
\begin{figure}
\epsfig{file=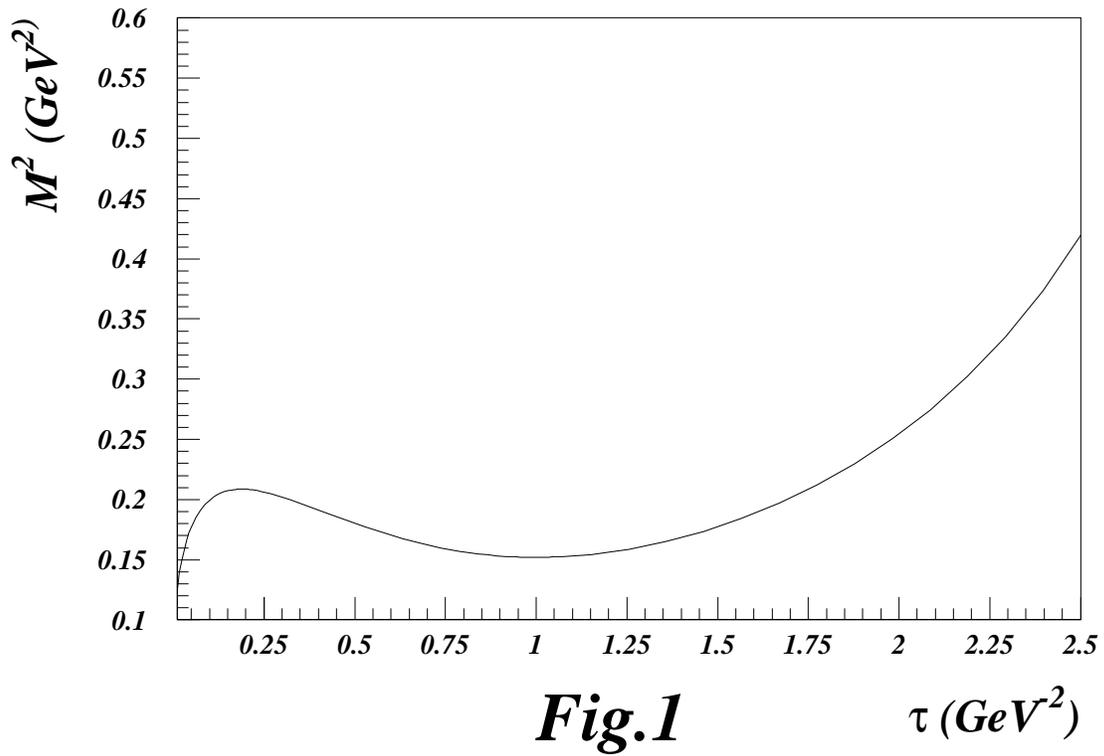}
\caption{$\frac{R_0}{R_{-1}}$ versus $\tau$ at $s_0=3.6$ GeV$^2$ without
radiative corrections.}
\end {figure}
\begin{figure}
\epsfig{file=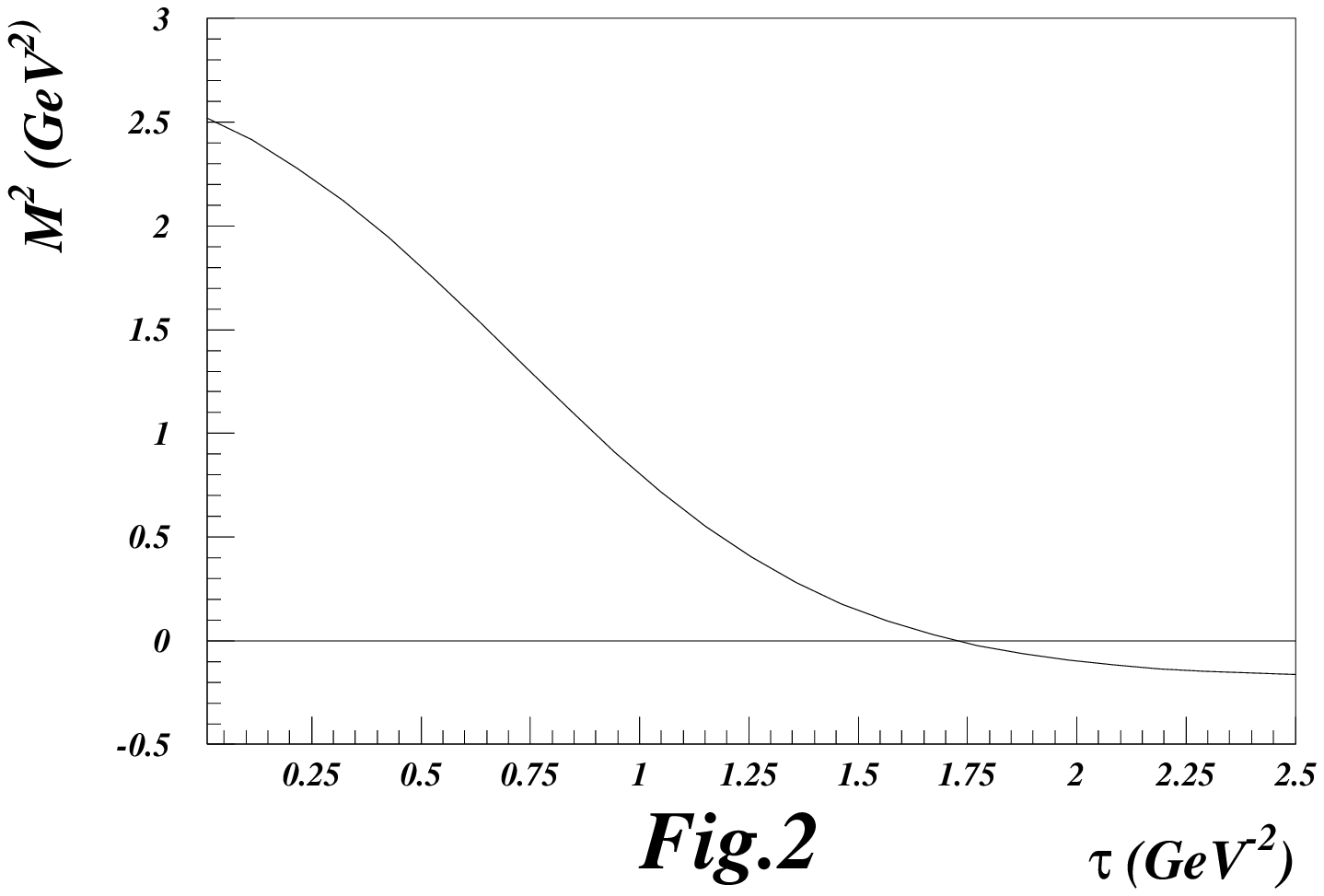}
\caption{$\frac{R_1}{R_0}$ versus $\tau$ at $s_0=3.6$ GeV$^2$ without
radiative corrections.}
\end {figure}
\begin{figure}
\epsfig{file=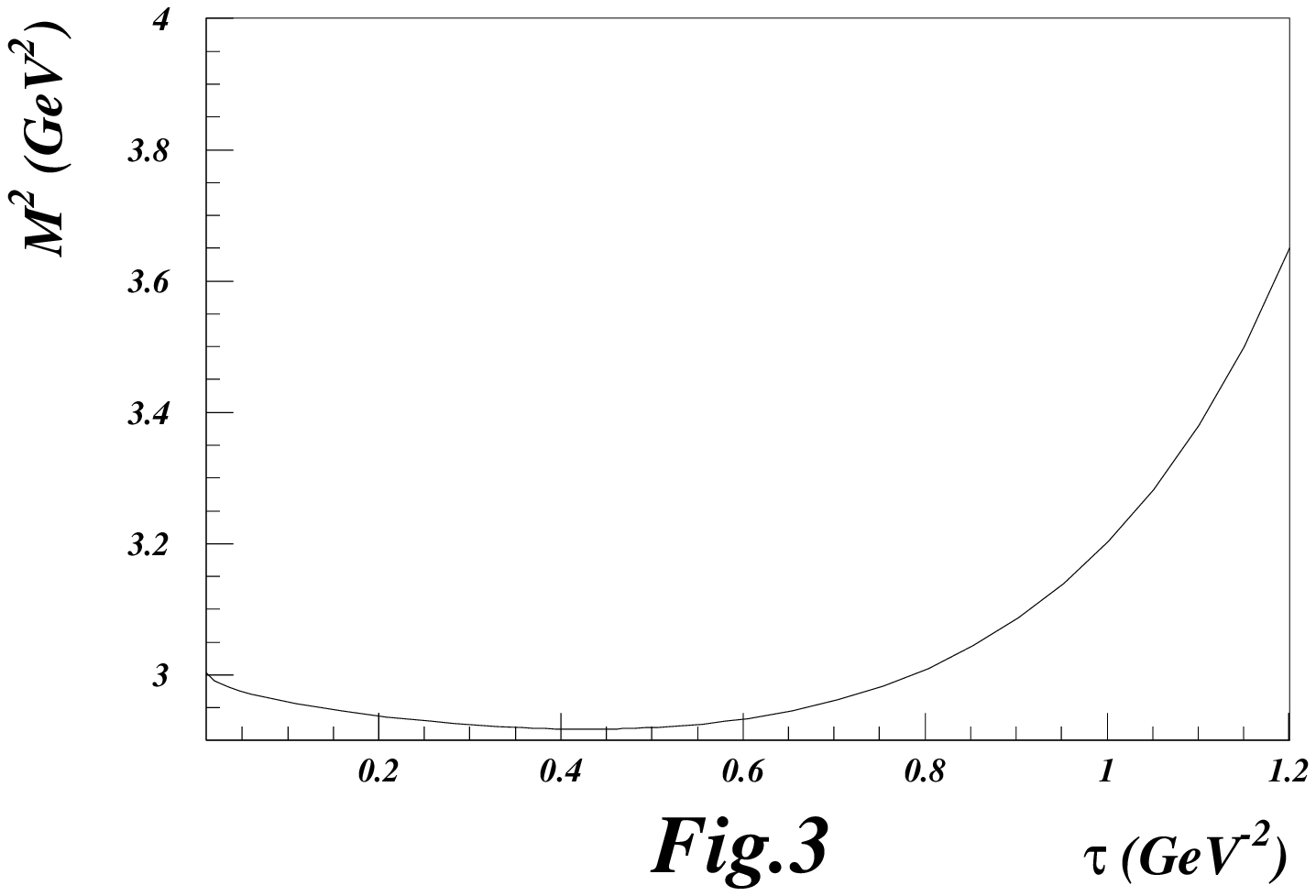}
\caption{$\frac{R_2}{R_1}$ versus $\tau$ at $s_0=3.6$ GeV$^2$ without
radiative corrections.}
\end {figure}
\begin{figure}
\epsfig{file=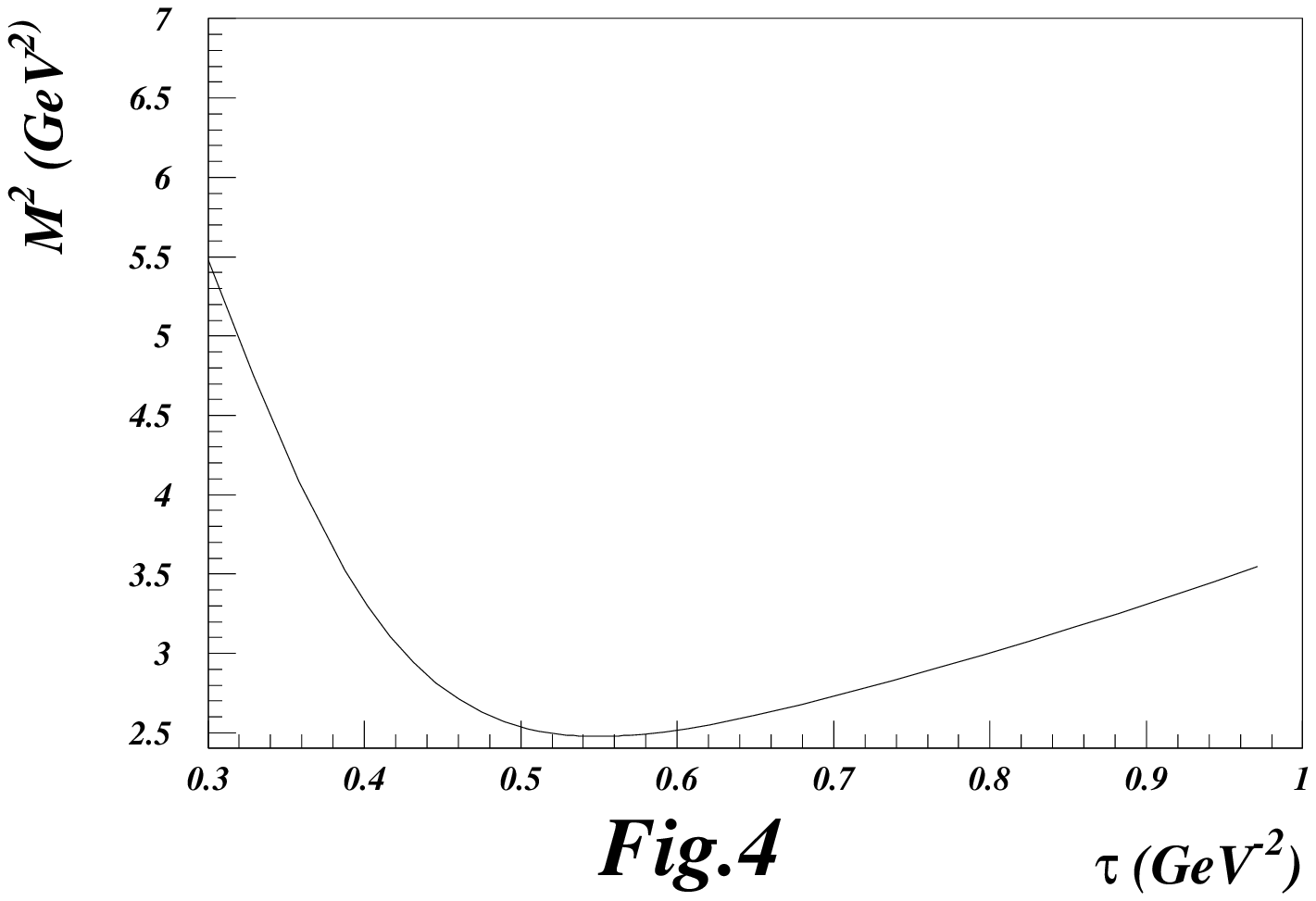}
\caption{$\frac{R_2}{R_1}$ versus $\tau$ at $s_0=3.6$ GeV$^2$ with radiative
corrections.}
\end {figure}
\end{document}